\documentclass[a4paper,11pt]{article}
\pdfoutput=1 % if your are submitting a pdflatex (i.e. if you have
\usepackage{jheppub} % for details on the use of the package, please
\usepackage[T1]{fontenc} % if needed
\usepackage{multirow}

\usepackage{graphicx}
\graphicspath{{figures/}{fig/}}
\usepackage{amsmath}
\usepackage{bm}
\usepackage{slashed}
\usepackage{epsfig}
\usepackage{amsfonts}
\usepackage{epstopdf}
\usepackage{color}
\usepackage{extarrows}
\allowdisplaybreaks
\newcommand{\etac}{{\eta_c}}

\newcommand{\state}[4]{{^{#1}\hspace{-0.6mm}#2_{#3}^{[#4]}}}

\newcommand\CSaSz{\state{1}{S}{0}{1}}

\def\qqb{{Q\overline{Q}}}
\def\SGFOetac{{\langle \overline{\cal O}_{\state{1}{S}{0}{1}}^{\eta_{c}} \rangle_{\text{ex}}}}
\def\NROetac{{\langle {\cal O}_{\state{1}{S}{0}{1}}^{\eta_{c}} \rangle_\text{ex}}}

\title{Exclusive quarkonium production or decay in soft gluon factorization}

\author[a]{Rong Li}
\author[b]{Yu Feng}
\author[c,d,e]{Yan-Qing Ma}

\affiliation[a]{School of Science, Xi'an Jiaotong University, Xi'an 710049,	China}
\affiliation[b]{Department of Physics, College of Basic Medical Sciences, Army Medical University, Chongqing 400038, China}
\affiliation[c]{School of Physics and State Key Laboratory of Nuclear Physics and Technology, Peking University,\\Beijing 100871, China}
\affiliation[d]{Center for High Energy Physics, Peking University,\\Beijing 100871, China}
\affiliation[e]{Collaborative Innovation Center of Quantum Matter,\\Beijing 100871, China}

\emailAdd{rongliphy@xjtu.edu.cn}
\emailAdd{yfeng@ihep.ac.cn}
\emailAdd{yqma@pku.edu.cn}

\abstract{
In this paper, we study the application of the recently proposed soft gluon factorization (SGF) to exclusive quarkonium production or decay. We find that in the nonrelativistic QCD factorization framework there are too many nonperturbative parameters. Thanks to the factorization of kinematical physics from dynamical physics, the SGF significantly reduces the number of nonperturbative parameters. Therefore, the SGF can improve our predictive power of  exclusive quarkonium production or decay. By applying to $\eta_c+\gamma$ production at B-factories, our result is the closest one to data among all theoretical calculations.
}

\begin{document}
	
\maketitle

\section{Introduction}

The study of heavy quarkonium production or decay is useful to understand both perturbative and nonperturbative physics of QCD. In the past two decades, the widely used theory for quarkonium physics is the non-relativistic quantum chromodynamics (NRQCD) factorization~\cite{Bodwin:1994jh}, which factorizes processes to perturbatively calculable hard parts multiplied by nonperturbative long-distance matrix elements (LDMEs). Because LDMEs are simply numbers, the NRQCD factorization has strong predictive power.

However, recent studies reveal that NRQCD factorization encounters some difficulties in describing inclusive quarkonium production data. In ref.~\cite{Ma:2017xno}, it was argued that the difficulties of NRQCD may be caused by the bad convergence of relativistic expansion. To overcome this problem, a
new factorization method called soft gluon factorization (SGF) was proposed~\cite{Ma:2017xno}. In this method, a series of relativistic corrections can be resummed to all orders and thus a better convergences in relativistic expansion is expected.

Exclusive quarkonium processes have also been paid a lot of attentions to. For example, the leading order (LO) in $\alpha_s$ calculation of $J/\psi+\eta_c$ production at B factories in NRQCD~\cite{Braaten:2002fi,Liu:2002wq} conflicts with experimental data~\cite{Abe:2002rb}, which attracts a lot of studies of relativistic corrections \cite{He:2007te,Bodwin:2007ga} and high order QCD corrections~\cite{Zhang:2005cha,Gong:2007db,Feng:2019zmt}. Later on, the factorization theorem for double charmonia production was proved rigorously~\cite{Bodwin:2008nf,Bodwin:2009cb,Bodwin:2010fi}.
The exclusive production of $\eta_c+\gamma$ at B factories has also been studied extensively within the NRQCD factorization~\cite{Chung:2008km,Li:2009ki,Sang:2009jc,Fan:2012dy,Li:2013nna,Xu:2014zra,Chen:2017pyi,Chung:2019ota}. Recently, this cross section was measured by the Belle Collaboration~\cite{Jia:2018xsy}, which found an unexpectedly small result comparing with all theoretical predictions.

Due to the advantage of SGF, it is interesting to see what happens if one applies it to exclusive quarkonium processes, which is the aim of this paper. In the rest of the paper, we first establish the general method to apply SGF on exclusive quarkonium processes. We then take $\sigma_{e^+e^- \to \eta_c+\gamma}$ and $\Gamma(\eta_c\to \gamma+\gamma)$ as examples to do phenomenological study. We will find that the SGF can not only factorize perturbative physics from nonperturbative physics but also factorize kinematical physics from dynamical physics. Because of the later effect, the SGF has much less free parameters comparing with NRQCD factorization, and therefore it has much stronger predictive power. Our result for  $\eta_c+\gamma$ production cross section is the closest one to data among all theoretical calculations.

\section{Factorization formula for exclusive processes}

\subsection{From inclusive formula to exclusive formula}
In SGF, the differential cross section for inclusive production of a quarkonium $H$ is given by~\cite{Ma:2017xno}
\begin{align}\label{eq:fac4d}
(2\pi)^3 2 P^{0} \frac{d\sigma_{H}}{d^3P}= \sum_{n} \int \frac{d^4 P_c}{(2\pi)^4} {\cal H}_{n}(P_c) F_{n\to H}(P_c,P),
\end{align}
where ${\cal H}_{n}(P_c)$ are perturbatively calculable hard parts that, roughly speaking, produce an intermediate state $n$ with total momentum $P_c$, and $F_{n\to H}(P_c,P)$ are called soft gluon distributions (SGDs) that describe the hadronization of the intermediate state to the quarkonium $H$ with momentum $P$.

Exclusive process can be thought of as a special case of inclusive process, where there is no real emission during the hadronization process, described by $F_{n\to H}(P_c,P)$. As a result, all conserved quantities are the same between the intermediate state $n$ and the quarkonium $H$, which include total momentum, color charge, and $J^{PC}$. The momentum conservation enables the decomposition
\begin{align}
F_{n\to H}(P_c,P)\to(2\pi)^4\delta^4(P_c-P) \langle \overline{\cal O}_n^{H} \rangle_{\text{ex}},
\end{align}
where the subscript ``ex'' means exclusive production. Then we get the factorization formula for exclusive  production of polarization summed quarkonium
\begin{align}\label{eq:facX}
(2\pi)^3 2 P^0 \frac{d\sigma_H}{d^3P}= \sum_{n}  {\cal H}_{n}(P) \langle \overline{\cal O}_n^{H} \rangle_{\text{ex}}.
\end{align}

\subsection{Definition of nonperturbative matrix elements}

As discussed in ref.~\cite{Ma:2017xno}, the intermediate state $n$ should at least contain a $\qqb$ pair. In addition to that, $n$ can also contain dynamical soft partons, i.e., gluons or light (anti-)quarks. Contributions from intermediate states with soft partons are inevitably suppressed by powers of $v$, the typical velocity of heavy quark inside of the quarkonium, and usually there is no mechanism to enhance this kind of contributions~\cite{Ma:2017xno}. Therefore, as the first approximation, we only consider intermediate states that do not contain dynamical soft partons and denote quantum numbers of $n$ in terms of spectroscopic notation $\state{{2S+1}}{L}{{J,J_z}}{1}$, with superscript $1$ denoting color singlet.
Then, we have
\begin{align}
\langle \overline{\cal O}_n^{H} \rangle_{\text{ex}}= \sum_{J_z^H} \langle 0| [\overline\Psi{\cal K}_{n} \Psi]^\dagger(0)|H\rangle_S \langle H| [\overline\Psi {\cal K}_{n}\Psi](0) |0\rangle_S,
\end{align}
where ``S'' means removing hard modes in the operator definition~\cite{Ma:2017xno}, the quarkonium state has standard relativistic normalization, and ${\cal K}_{n}$ are defined as~\cite{Ma:2017xno}
\begin{align}\label{eq:Kn}
{\cal K}_n=\frac{\sqrt{M}}{M+2m}\frac{M+\slashed{P}}{2M}\Gamma_{n}\frac{M-\slashed{P}}{2M}\,  {\cal C}^{[1]}\,,
\end{align}
where $m$ is the mass of heavy quark, $M$ is the mass of heavy quarkonium, and  color projector ${\cal C}^{[1]}=1/\sqrt{N_c}$. If $n$ is S-wave, we have
\begin{align}\label{eq:spinProj}
	\Gamma_{n}=\begin{cases}
	\gamma_5,~~~~~~~&\text{if $S=0$,}\\
	{\epsilon}_{S_z}^\mu \gamma_\mu,~~&\text{if $S=1$,}
	\end{cases}
\end{align}
where ${\epsilon}_{S_z}^\mu$ is polarization vector with $P\cdot {\epsilon}_{S_z}=0$. For other cases, we refer the definition of $\Gamma_{n}$ in ref.~\cite{Ma:2017xno}.
Following the discussion in ref.~\cite{Ma:2017xno}, we can obtain
\begin{align}\label{eq:relation}
\langle \overline{\cal O}_n^{H} \rangle_{\text{ex}}= \frac{1}{2N_c}\langle {\cal O}_n^{H} \rangle_{\text{ex}} [1+O(v^2)],
\end{align}
where $\langle {\cal O}_n^{H} \rangle_{\text{ex}}$ are standard NRQCD LDMEs for exclusive quarkonium production, which can be related to inclusive quarkonium production LDMEs via vacuum saturation approximation \cite{Bodwin:1994jh},
\begin{align}
\langle {\cal O}_n^{H} \rangle_{\text{ex}} =\langle {\cal O}_n^{H} \rangle [1+O(v^4)].
\end{align}

\subsection{Factorization formula at the amplitude level}

Based on the above definition, it is clear that the factorization formula eq.~\eqref{eq:facX} can be expressed at the amplitude level as \footnote{Considered the fact that different intermediate states can interfere with each others, the factorization formula at the amplitude level is more precise.}
\begin{align}\label{eq:facA}
\mathcal{A}_{H}(P)=\sum_{n} \hat{\mathcal{A}}_{n}(P) \overline{R}^n_H,
\end{align}
where
\begin{align}\label{eq:Rt}
\overline{R}^n_H= \langle H|[\overline\Psi {\cal K}_{n}\Psi](0) |0\rangle_S.
\end{align}
with
\begin{align}
\langle \overline{\cal O}_n^{H} \rangle_{\text{ex}}=&\sum_{J_z^H} \left|\overline{R}^n_H\right|^2,\\
\sigma_H=&\frac{1}{2s}\int \overline{\sum} |\mathcal{A}_{H}(P)|^2 d\text{PS},
\end{align}
where ``$\overline{\Sigma}$'' means averaging over initial states and summing over final states.  Note that, based on the relation eq.~\eqref{eq:relation}, $\overline{R}^n_H$ can be related to wave function at the origin of the heavy quarkonium \cite{Bodwin:1994jh} up to $O(v^2)$ corrections.

\subsection{Perturbative matching}

The factorization formula eq.~\eqref{eq:facA} will be our start point to study exclusive quarkonium production in the following. To use this formula, we need to calculate $\hat{\mathcal{A}}_{n}(P)$ in perturbation theory. To this end, we relable $n$ in eq.~\eqref{eq:facA} by $n'$ and then project $H$ to a color-singlet state $n=\qqb\left(\state{{2S+1}}{{L}}{{J}}{1}\right)$ in both sides of the equation, which results in
\begin{align}\label{eq:facPer}
\mathcal{A}_{n}(P)=\sum_{n'} \hat{\mathcal{A}}_{n'}(P) \overline{R}^{n'}_{n},
\end{align}
where the definition of projection will be explained below.

Following ref.~\cite{Ma:2017xno}, we define a complete set of on-shell color-singlet state $n=\qqb\left(\state{{2S+1}}{{L}}{{J}}{1}\right)$ with momenta
\begin{align}
p_Q&=P/2+q,\\
p_{\overline{Q}}&=P/2-q.
\end{align}
We project the pair to color-singlet state by ${\cal C}^{[1]}$, and we project the pair to state with total spin $S$ and $S_z$ by replacing spinors of $\qqb$ pair by
\begin{align}
\widetilde{\Pi}_{SS_z}&=\frac{(\slashed{p}_{\overline{Q}}-m) \frac{M-\slashed{P}}{2M} \widetilde{\Gamma}^s_{SS_z} \frac{M+\slashed{P}}{2M} (\slashed{p}_{{Q}}+m)}{\sqrt{M}(M/2+m)},
\end{align}
with
\begin{subequations}\label{eq:spinProj}
	\begin{align}
	\widetilde{\Gamma}^s_{00} &= -\gamma_5,\\
	\widetilde{\Gamma}^s_{1S_z} &= {\epsilon}_{S_z}^{*\mu} \gamma_\mu.
	\end{align}
\end{subequations}
On-shell conditions $p_Q^2=p_{\bar Q}^2=m^2$ results in
\begin{align}
&P\cdot q=0,\\
&q^2=m^2-M^2/4,
\end{align}
which constrain two degrees of freedom of $q$. The other two degrees of freedom of $q$, defined as spatial angles in the rest frame of the pair, can be removed by partial wave expansion.

Eventually, we get our definition
\begin{align}\label{eq:An}
&\mathcal{A}_{n}(P)=\sum_{L_z,S_z} \langle L, L_z; S,S_z |J,J_z \rangle \int d^2 \Omega  |{\bm q}|^{-L} \sqrt{\frac{(2L+1)!!}{4\pi (L!)}} Y_{L}^{*L_z} \text{Tr}\left[\mathcal{C}^{[1]} \widetilde{\Pi}_{SS_z} \mathcal{A}_{Q+\overline{Q}}(P)\right]\,,
\end{align}
where $\mathcal{A}_{Q+\overline{Q}}(P)$ is the $\qqb$ production amplitude with spinors of $\qqb$ removed.
Similarly,
\begin{align}
\overline{R}^{n'}_{n}&=\sum_{L_z,S_z} \langle L, L_z; S,S_z |J,J_z \rangle \int d^2 \Omega  |{\bm q}|^{-L}\sqrt{\frac{(2L+1)!!}{4\pi (L!)}} Y_{L}^{*L_z} \text{Tr}\left[\mathcal{C}^{[1]} \widetilde{\Pi}_{SS_z} \overline{R}^{n'}_{Q+\overline{Q}}\right]\,.
\end{align}
Following the derivation in ref.~\cite{Ma:2017xno}, we have
\begin{align}\label{eq:Rn}
\overline{R}^{n'(0)}_{n}&=\delta_{nn'}\,,
\end{align}
where the superscript ``(0)'' denotes leading order in $\alpha_s$ expansion.

By inserting perturbative expansions
\begin{subequations}
	\begin{align}
	\mathcal{A}_{n}&=\mathcal{A}_{n}^{(0)}+\alpha_s \mathcal{A}_{n}^{(1)}+\alpha_s^2 \mathcal{A}_{n}^{(2)}+\cdots,\\
	\hat{\mathcal{A}}_{n'}&=\hat{\mathcal{A}}_{n'}^{(0)}+\alpha_s \hat{\mathcal{A}}_{n'}^{(1)}+\alpha_s^2 \hat{\mathcal{A}}_{n'}^{(2)}+\cdots,\\
	\overline{R}^{n'}_{n}&=\overline{R}^{n'(0)}_{n}+\alpha_s \overline{R}^{n'(1)}_{n}+\alpha_s^2 \overline{R}^{n'(2)}_{n}+\cdots
	\end{align}
\end{subequations}
into eq.~\eqref{eq:facPer} and using the orthogonal relations eq.~\eqref{eq:Rn}, we get the following relations
\begin{subequations}\label{eq:match}
	\begin{align}
	\hat{\mathcal{A}}_{n}^{(0)}=&\mathcal{A}_{n}^{(0)},\\
	\begin{split}
	\hat{\mathcal{A}}_{n}^{(1)}=&\mathcal{A}_{n}^{(1)}-\sum_{n^\prime} \mathcal{A}_{n^\prime}^{(0)}\overline{R}^{n'(1)}_{n},
	\end{split}\\
	\begin{split}
	\hat{\mathcal{A}}_{n}^{(2)}=&\mathcal{A}_{n}^{(2)}-\sum_{n^\prime} \mathcal{A}_{n^\prime}^{(1)} \overline{R}^{n'(1)}_{n}-\sum_{n^\prime} \mathcal{A}_{n^\prime}^{(0)} \overline{R}^{n'(2)}_{n},
	\end{split}
	\end{align}
\end{subequations}
and so on. Based on these relations, to get the perturbative expansion of $\hat{\mathcal{A}}_{n}$, we need to calculate $\mathcal{A}_{n}$ and
$\overline{R}_{n}^{n'}$ perturbatively.

\subsection{Factorization formula for quarkonium exclusive decay}

For quarkonium exclusive decay, we have similar factorization formula at the amplitude level
\begin{align}\label{eq:facGA}
\mathcal{A}^{H}=\sum_{n} \hat{\mathcal{A}}^{n} \overline{R}^{n*}_{H},
\end{align}
with decay width
\begin{align}
\Gamma^H=&\frac{1}{2M}\int \overline{\sum}|\mathcal{A}^{H}|^2 d\text{PS}.
\end{align}

Similar to quarkonium production, $\hat{\mathcal{A}}^{n}$ can be perturbatively calculated with
\begin{subequations}\label{eq:match}
	\begin{align}
	\hat{\mathcal{A}}^{n(0)}=&\mathcal{A}^{n(0)},\\
	\begin{split}
	\hat{\mathcal{A}}^{n(1)}=&\mathcal{A}^{n(1)}-\sum_{n^\prime} \mathcal{A}^{n'(0)}\overline{R}^{n'*(1)}_{n},
	\end{split}
	\end{align}
\end{subequations}
and so on. In these relations, ${\mathcal{A}}^{n}$ is defined by
\begin{align}
&\mathcal{A}^{n}=\sum_{L_z,S_z} \langle L, L_z; S,S_z |J,J_z \rangle \int d^2 \Omega  |{\bm q}|^{-L} \sqrt{\frac{(2L+1)!!}{4\pi (L!)}} Y_{L}^{L_z} \text{Tr}\left[\mathcal{C}^{[1]} {\Pi}_{SS_z} \mathcal{A}^{Q+\overline{Q}}\right]\,,
\end{align}
where $\mathcal{A}^{Q+\overline{Q}}$ is the $\qqb$ decay amplitude with spinors of $\qqb$ removed and
\begin{align}
{\Pi}_{SS_z}&=\frac{(\slashed{p}_{{Q}}+m) \frac{M+\slashed{P}}{2M} {\Gamma}^s_{SS_z} \frac{M-\slashed{P}}{2M} (\slashed{p}_{\overline{Q}}-m)}{\sqrt{M}(M/2+m)},
\end{align}
with
\begin{subequations}
	\begin{align}
	{\Gamma}^s_{00} &= - \gamma_5,\\
	{\Gamma}^s_{1S_z} &= {\epsilon}_{S_z}^\mu \gamma_\mu.
	\end{align}
\end{subequations}

\section{Applications}

\subsection{Leading order calculation of $e^+e^- \to \etac+\gamma$ }

For $\eta_c$ production, we have $J^{PC}=0^{-+}$, which demands $n=\CSaSz$. Thus the amplitude of $e^+e^- \to \etac+\gamma$ is given by
\begin{align}\label{eq:fac}
\begin{split}
\mathcal{A}_{e^+e^- \to \eta_c+\gamma}
= \hat{\mathcal{A}}_{e^+e^- \to c\bar{c}(\CSaSz)+\gamma}\overline{R}^{\CSaSz}_{\eta_c}.
\end{split}
\end{align}
At the lowest order in $\alpha_s$, we have
\begin{align}
\begin{split}
\hat{\mathcal{A}}_{e^+e^- \to c\bar{c}(\CSaSz)+\gamma}^{(0)}=\mathcal{A}_{e^+e^- \to c\bar{c}(\CSaSz)+\gamma}^{(0)},
\end{split}
\end{align}
with
\begin{align}\label{eq:amp}
\begin{split}
&\mathcal{A}_{e^+e^- \to c\bar{c}(\CSaSz)+\gamma}^{(0)}=\int \frac{d^2\Omega}{4\pi} \text{Tr}\left[\mathcal{C}^{[1]} \widetilde{\Pi}_{00} \mathcal{A}_{e^+e^- \to c+\bar{c}+\gamma}^{(0)}\right].
\end{split}
\end{align}

\begin{figure}[htb!]
 \begin{center}
 \includegraphics[width=0.9\textwidth]{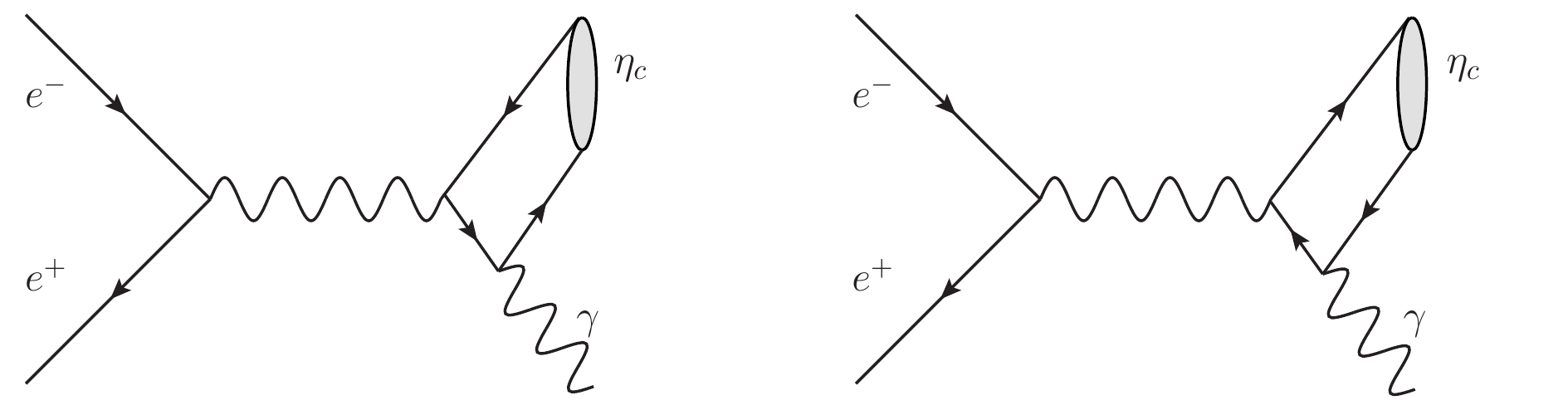}
  \caption{ Feynman diagrams for the process $e^+e^- \to \etac+\gamma$ at leading order in $\alpha_s$ . \label{fig:diags}}
 \end{center}
\end{figure}
Therefore, we need to calculate the two Feynman diagrams in figure.~\ref{fig:diags}. Since the leptonic current is independent of spatial angles, we separate it from
the complete amplitude and rewrite eq.~\eqref{eq:amp} as
\begin{align}\label{eq:qqbar}
\begin{split}
%\mathcal{A}_{e^+e^- \to \eta_c+\gamma}(P)=\mathcal{L}_{\mu} \mathcal{A}^{\mu} \\
\mathcal{A}^{(0)}_{e^+e^- \to c\bar{c}(\CSaSz)+\gamma}=\frac{-i}{s}\mathcal{L}_{\mu} \epsilon^{\ast}_{\nu}(k)\mathcal{A}^{\mu\nu}
\end{split}
\end{align}
where $s=(k_{e^+}+k_{e^-})^2$ is the square of center of mass energy, $\epsilon^{\ast}_{\nu}(k)$ is the polarization vector of the final-state photon
and the leptonic current $\mathcal{L}_{\mu}$ is defined by
\begin{align}
\mathcal{L}_{\mu}=-ie\,\overline{v}(k_{e^+})\gamma_{\mu} u(k_{e^-}).
\end{align}

Lorentz symmetry, C, P and T invariance imply that the hadronic current can be expressed as
\begin{align}\label{eq:r2qq}
\begin{split}
\mathcal{A}^{\mu\nu}= \epsilon^{\mu\nu Pk} A,
\end{split}
\end{align}
where A is a lorentz invariant quantity.
Then we multiply $\epsilon_{\mu\nu Pk}$ on both sides of eq.~\eqref{eq:r2qq} and sum over Lorentz indexes, we obtain
\begin{align}
\begin{split}
A=\frac{\mathcal{A}^{\mu\nu} \epsilon_{\mu\nu Pk}}{2(P\cdot k)^2},
\end{split}
\end{align}
where we have used $\epsilon^{\mu\nu Pk}\epsilon_{\mu\nu Pk}=2(P\cdot k)^2$.
Then the integration over spatial angles is very easy, which eventually gives
\begin{align}
\begin{split}
A=\frac{96\pi\alpha\, m\, e_c^2}{M\sqrt{3M\delta}(s-M^2)}\ln(\frac{1+\sqrt{\delta}}{1-\sqrt{\delta}})
\end{split}
\end{align}
where $\delta=1-4m^2/M^2$ and electronic charge of charm quark is denoted by $e_c \, e$.

By substituting for leptonic current
\begin{align}
\begin{split}
\mathcal{L}_{\mu}\mathcal{L}^{\ast}_{\mu'}\to-\frac{4e^2s}{3}g_{\mu\mu'}
\end{split}
\end{align}
and for the summation of photon polarization
\begin{align}
\begin{split}
\sum\epsilon^{\ast}(k)_{\nu}\epsilon(k)_{\nu'}\to-g_{\nu\nu'},
\end{split}
\end{align}
we get the cross section
\begin{align}\label{eq:xsection}
\begin{split}
\sigma_{e^+e^- \to \eta_c+\gamma}=&\frac{1}{2s}\int\overline{|\mathcal{A}_{e^+e^- \to \eta_c+\gamma}|^2}d\text{PS}_2 \\
=&\frac{1}{2s}\int(\frac{1}{s^2})(\frac{1}{4})(-\frac{4e^2s}{3}g_{\mu\mu'})(-g_{\nu\nu'})\epsilon^{\mu\nu Pk}\epsilon^{\mu'\nu' Pk}  |A|^2 \SGFOetac d\text{PS}_2 \\
=&\frac{128\pi^2\alpha^3 m^2 e_c^4(s-M^2)}{s^3 M^3 \delta }
\ln^2(\frac{1+\sqrt{\delta}}{1-\sqrt{\delta}})\SGFOetac\,.
%%&\ln^2(\frac{1+\sqrt{\delta}}{1-\sqrt{\delta}})|\, R_{\CSaSz}^{\eta_Q}|^2
\end{split}
\end{align}

According to the proposal in SGF~\cite{Ma:2017xno}, one can expand $m$ in the hard part around $M/2$ to simply the calculation, which is equivalent to expand eq.~\eqref{eq:xsection} in power series of $\delta$, which gives
\begin{align}\label{eq:sigma}
\begin{split}
\sigma_{e^+e^- \to \eta_c+\gamma}=&\frac{128\pi^2\alpha^3 e_c^4(s-M^2)}{s^3 M }
\SGFOetac\left(1-\frac{1}{3}\delta-\frac{7}{45}\delta^2-\frac{29}{315}\delta^3+ ...\right),
\end{split}
\end{align}
where high order terms in this expansion can be thought of as relativistic corrections to the lowest order term.

\subsection{Leading order calculation of $\etac \to \gamma \gamma$ }

We begin with the amplitude of $\etac \to \gamma \gamma$
\begin{align}\label{eq:dec1}
\begin{split}
\mathcal{A}^{\etac \to \gamma\gamma}
= \hat{\mathcal{A}}^{c\bar{c}(\CSaSz)\to \gamma\gamma}\overline{R}^{\CSaSz*}_{\eta_c},
\end{split}
\end{align}
with
\begin{align}\label{eq:amp-decay}
\begin{split}
\mathcal{A}^{c\bar{c}(\CSaSz)\to \gamma\gamma}
= \int d\Omega \text{Tr}\left[\mathcal{C}^{[1]} \Pi_{00} \mathcal{A}^{{c+\bar{c}\to \gamma\gamma}}\right].
\end{split}
\end{align}
The amplitude can be decomposed as
\begin{align}\label{eq:qqbar-decay}
\begin{split}
\mathcal{A}^{c\bar{c}(\CSaSz)\to \gamma\gamma}=\epsilon^{\ast}_{\mu}(k_1)\epsilon^{\ast}_{\nu}(k_2)\mathcal{A}^{\mu\nu},
\end{split}
\end{align}
where $k_i(i=1,2)$ are momenta of  $\gamma$'s in the final state and the hadronic current $\mathcal{A}^{\mu\nu}$
can be calculated similarly to the production case.

The decay width of $\eta_c \to \gamma\gamma$ in SGF gives
\begin{align}\label{eq:Getac}
\begin{split}
\Gamma(\eta_c \to \gamma\gamma)=&\frac{1}{2M}\cdot\frac{1}{2!}\int\overline{|\mathcal{A}_{\etac \to \gamma\gamma}|^2}d\text{PS}_2\\
=&\frac{1}{2M}\cdot\frac{1}{2!}\cdot (-g_{\mu\mu'})(-g_{\nu\nu'})\epsilon^{\mu\nu Pk}\epsilon^{\mu'\nu' Pk}  |A|^2 \frac{1}{8\pi} \SGFOetac \\
=&\frac{48\pi\alpha^2 m^2 e_c^4}{M^4\delta}
\ln^2(\frac{1+\sqrt{\delta}}{1-\sqrt{\delta}})\SGFOetac.
\end{split}
\end{align}
This expression is consistent with NRQCD calculation in ref.~\cite{Bodwin:2007fz} and references therein, although the meaning of $\delta$ is significantly different in SGF and that in NRQCD. In the former case, $\delta$ is a constant in perturbative hard part, while in the later case $\delta$ is a {\it free} parameter in QCD calculation and Taylor expansion by which defines perturbative hard part.

\subsection{Numerical result}

Based on eqs.~\eqref{eq:xsection} and \eqref{eq:Getac}, we get
\begin{align}\label{eq:xs2wd}
\begin{split}
\sigma_{e^+e^- \to \eta_c+\gamma}=\frac{8\pi\alpha M(s-M^2)}{3s^3}\Gamma(\eta_c \to \gamma\gamma),
\end{split}
\end{align}
which relates the exclusive production cross section of $\eta_c$ with the exclusive decay width. This relation holds at leading order in $\alpha_s$ and all orders in $v$ if we ignore contributions from operators with explicit dynamical soft fields. It is impressive that this relation has no dependence on heavy quark mass $m$.
We note that a relation similar to eq. \eqref{eq:xs2wd} also exists in NRQCD, which can be obtained by simply replacing $M$ by $2 m_c$ in eq. \eqref{eq:xs2wd}. However, the similar relation in NRQCD is valid only at leading order in $v$ and it depends on heavy quark mass. As a result, the prediction for $\sigma_{e^+e^- \to \eta_c+\gamma}$ using eq. \eqref{eq:xs2wd} in SGF has almost no free parameters, while the prediction using the similar relation in NRQCD depends on how to play with heavy quark mass and relativistic corrections.

In numerical calculation, we choose the fine structure constant $\alpha=1/137$, the quarkonium mass $M=2.98$ GeV and collision energy $\sqrt{s}=10.58$ GeV for B factories.
By using experiment result for decay width~\cite{Tanabashi:2018oca}
\begin{align}
\Gamma(\eta_c \to \gamma\gamma)&=5.02\pm 0.13\pm 0.38 \text{ keV},
\end{align}
we get the prediction for cross section of $e^+e^- \to \eta_c+\gamma$ in SGF:
\begin{align}
\sigma_{e^+e^- \to \eta_c+\gamma}=26.2 \pm 2.1 \text{ fb} .
\end{align}
Note that here we only include uncertainties from experimental side, but have not considered theoretical ones. Considering $\alpha_s$ corrections and $v^2$ corrections from operators with explicit dynamical soft fields, a $30\%$ uncertainty should be expected, which amounts to $\pm 8$ fb.

If we further choose the heavy quark pole mass as $m=1.4\pm0.2$ GeV, we obtain
\begin{align} \label{eq:Oetac}
2N_c\SGFOetac=0.176\pm 0.014^{+0.021}_{-0.015}=0.176^{+0.025}_{-0.021}\text{GeV}^3.
\end{align}
where the later uncertainty is due to the varying of heavy quark mass.

\section{Comparison with NRQCD factorization and the Belle's measurement}

In the NRQCD factorization~\cite{Bodwin:1994jh}, the calculation of the cross section of $e^+e^- \to \eta_c+\gamma$ is very similar to that in the SGF method, but one needs to expand $q^2$ in the hard part and then put it into the definition of LDMEs. By setting $M^2=4m^2-4q^2$, from eq.~\eqref{eq:xsection} we can obtain the cross section calculated in NRQCD:
\begin{align}\label{eq:Xnr}
\begin{split}
\sigma_{e^+e^- \to \eta_c+\gamma}=&\hat{\sigma}_{\CSaSz}^{v^0} \langle \mathcal{O}^{\eta_{c}}(\CSaSz)\rangle_\text{ex} +\hat{\sigma}_{\CSaSz}^{v^2} \langle \mathcal{P}^{\eta_{c}}(\CSaSz)\rangle_\text{ex}+\cdots,
\end{split}
\end{align}
with short-distance coefficients
\begin{align}
\hat{\sigma}_{\CSaSz}^{v^0}&=\frac{32\pi^2\alpha^3e_c^4(s-4m^2)}{3m s^3},\\
\hat{\sigma}_{\CSaSz}^{v^2}&=-\frac{16\pi^2\alpha^3e_c^4(5s+4m^2)}{9m s^3},
\end{align}
where $\hat{\sigma}_{\CSaSz}^{v^0}$ agrees with results in refs.~\cite{Li:2009ki,Braguta:2010mf} and $\hat{\sigma}_{\CSaSz}^{v^2}$ agrees with results in refs.~\cite{Li:2009ki,Sang:2009jc,Fan:2012dy,Li:2013nna,Xu:2014zra}.

The terms in eq.~\eqref{eq:Xnr} are organized by the power counting in $v$, e.g., the second term is $v^2$ suppressed comparing with the first term. As each term introduces one free parameter, in NRQCD factorization one needs a lot of nonperturbative parameters to provide a precise description of the cross section. In contrast, one needs only one parameter in SGF framework \footnote{Note that, both in NRQCD and SGF, we have ignored intermediate states that explicitly include dynamical soft partons.}.

In table~\ref{tab:N-results}, we compare predictions based on SGF and that based on NRQCD factorization~\cite{Sang:2009jc,Li:2013nna,Fan:2012dy}.  Because of different choice of parameters, results at lowest order in $v$ are very different between these references. Especially, the $\NROetac$ is chosen as $0.437 \text{GeV}^3$ in refs.~\cite{Sang:2009jc,Fan:2012dy} and  $0.694 \text{GeV}^3$ in ref.~\cite{Li:2013nna}. The difference of the parameters can be well understood. For example, comparing our result in eq.~\eqref{eq:Oetac} with $\NROetac=0.437^{+0.111}_{-0.105} \text{GeV}^3$ obtained in ref. \cite{Bodwin:2007fz}, nearly half of the suppression is due to the change of experimental value $\Gamma(\eta_c \to \gamma\gamma)$ from the older one $7.2\pm 0.7\pm2.0 \text{ keV}$ to the new one $5.02\pm 0.13\pm 0.38 \text{ keV}$. About another half of the suppression is due to the missing of $\alpha_s$ correction in our result. As we have not calculated $\alpha_s$ corrections, the value in eq.~\eqref{eq:Oetac} can be only consistently used for leading order in $\alpha_s$ calculations, like that in eq.~\eqref{eq:sigma}.

To improve theoretical precision in NRQCD factorization, in ref.~\cite{Fan:2012dy} the authors resum the relativistic correction to all orders for $e^+e^- \to \eta_c+\gamma$. This resummation in principle needs infinity number of nonperturbative parameters, which are modeled by assuming a generalized Gremm-Kapustin relation \cite{Bodwin:2006dm} that relates all LDMEs at higher order in $v$ to $\langle \mathcal{O}^{\eta_{c}}(\CSaSz)\rangle_\text{ex}$. The resummed result is also listed in table~\ref{tab:N-results}. One can find that the SGF prediction is smaller than all predictions using NRQCD.

%\begin{table}
%\caption[]{The cross section $\sigma_{e^+e^- \to \eta_c+\gamma}$ calculated in SGF and NRQCD. }
%\label{tab:N-results}
%\renewcommand{\arraystretch}{1.5}
%\[
%\begin{array}{|c|c|c|c|c|c|}
%\hline \hline  &\textrm{SGF}&\textrm{NRQCD~\cite{Sang:2009jc}}&\textrm{NRQCD~\cite{Li:2013nna}}& \textrm{NRQCD~\cite{Fan:2012dy}}& \textrm{Ex.~\cite{}}\\
%\hline \textrm{$v^0$}&&83.3&117&83.2&\\
%\hline \textrm{$v^2$}&&73.5&92&&\\
%\hline \textrm{$v^{\infty\star}$}&&&&68.9&\\
%\hline \textrm{$v^{\infty}$}&26.2&&&&\\
%\hline \hline
%\end{array}
%\]
%\renewcommand{\arraystretch}{1.0}
%\end{table}

\begin{table}[h]
\normalsize
\centering
\caption[]{The cross section $\sigma_{e^+e^- \to \eta_c+\gamma}$ calculated in SGF and NRQCD at LO in $\alpha_s$, as well as the experimental upper limit  measured by the Belle Collaboration (in unit of fb).}
\label{tab:N-results}
\begin{tabular}{|c|c|c|c|c|c|}
\hline\hline  &\textrm{SGF}&\textrm{NRQCD~\cite{Sang:2009jc}}&\textrm{NRQCD~\cite{Li:2013nna}}& \textrm{NRQCD~\cite{Fan:2012dy}}& \textrm{Ex.$^{\text{UL}}$~\cite{Jia:2018xsy}} \\ \hline
\textrm{$v^0$}&&83.3&117&83.2&  \multirow{4}{*}{21.1} \\ \cline{1-5}
\textrm{$v^2$}&&73.5&92&&  \\ \cline{1-5}
\textrm{$v^{\infty\star}$}&&&&68.9& \\ \cline{1-5}
\textrm{$v^{\infty}$}&26.2&&&& \\ \hline\hline
\end{tabular}
\end{table}

It is interesting to understand why there are less parameters in SGF comparing with NRQCD. Let us begin with the Gremm-Kapustin relation \cite{Gremm:1997dq},
\begin{align}
\langle \mathcal{P}^{\eta_{c}}(\CSaSz)\rangle_{\text{ex}}= \frac{1}{2}m (M-2m) \langle \mathcal{O}^{\eta_{c}}(\CSaSz)\rangle_\text{ex} (1+O(v^2)),
\end{align}
where $O(v^2)$ terms are caused by matrix elements of operators with dynamical gluon fields. The Gremm-Kapustin relation tells us that nonperturbative behavior (infrared divergences in perturbation theory) of $\langle \mathcal{P}^{\eta_{c}}(\CSaSz)\rangle_{\text{ex}}$ can be decomposed by that of leading LDME $\langle \mathcal{O}^{\eta_{c}}(\CSaSz)\rangle_\text{ex}$ and other LDMEs with dynamical gluon fields. Therefore, for the purpose of a valid factorization, the introduction of $\langle \mathcal{P}^{\eta_{c}}(\CSaSz)\rangle_{\text{ex}}$ into the factorization formula is unnecessary.

In NRQCD factorization, one introduces nonperturbative quantities like $\langle \mathcal{P}^{\eta_{c}}(\CSaSz)\rangle_{\text{ex}}$ into factorization formula so that hard parts in eq.~\eqref{eq:Xnr} are independent of quarkonium mass. But the price to pay is that one has lots of nonperturbative parameters.

In SGF, we do not introduce nonperturbative quantities similar to $\langle \mathcal{P}^{\eta_{c}}(\CSaSz)\rangle_{\text{ex}}$ (see discussion in Appendix A of ref.~\cite{Ma:2017xno}). Leading contributions of these quantities are purely kinematic and they have been taken into account by coefficients of $\SGFOetac$.  Other contributions can be taken into account by nonperturbative SGDs with dynamical gluon fields, which are neglected in this paper but can be systematically included. In this sense, besides factorizing perturbative physics from nonperturbative physics, the SGF also factorizes kinematical physics from dynamical physics. It is the later effect that enables SGF to have less free parameters and thus to have a stronger predictive power.

As for the experimental aspect, recently Belle Collaboration measured the total cross section for $e^+e^- \to \eta_c+\gamma$ at various center of mass energies~\cite{Jia:2018xsy} and gave the upper limit for the total cross section of this process as $\sigma^{\text{UL}}=21.1 \text{fb}$ with $\sqrt{s}=10.58 \text{GeV}$ at 90\% credibility level. It can be seen from table~\ref{tab:N-results} that the SGF result is closest to the upper limit of experimental result. Considering also theoretical uncertainties from higher order corrections, our result is consistent with the Belle's measurement.

\section{Summary and outlook}

In summary, we set up a general framework to apply the newly proposed SGF to exclusive quarkonium production or decay. Comparing with the NRQCD factorization, our method resums a series of relativistic corrections to all orders, which can reduce theoretical uncertainties. More importantly, our method has much less number of nonperturbative parameters and therefore has a stronger predictive power. To make it possible, the SGF not only factorizes perturbative physics from nonperturbative physics but also factorizes kinematical physics from dynamical physics. Although corrections from kinematical physics and dynamical physics have similar velocity power counting in NRQCD effective theory, they may be significantly different in size for phenomenological problems.  Usually, we expect the kinematical part may have more important contributions. Therefore, the SGF, which resums the series of relativistic corrections originating from kinematical effects, may have a better convergence of velocity expansion comparing with NRQCD.

Taking $\sigma_{e^+e^- \to \eta_c+\gamma}$ and $\Gamma(\eta_c\to \gamma+\gamma)$ as examples, we show how to use our new method to do phenomenological study. We find that these two quantities can be related to each other in eq. \eqref{eq:xs2wd}, which holds to all orders in $v$, if we ignore operators involving explicit dynamical soft fields, and is independent of charm quark mass. Based on experimental inputs, we provide a prediction for $\sigma_{e^+e^- \to \eta_c+\gamma}$. Comparing with other predictions, our result agrees best with Belle's measurement.

%--------------------------------------------------------------------
\begin{acknowledgments}
%--------------------------------------------------------------------

We thank An-Ping Chen for many useful discussions. The work is supported in part by the National Natural Science Foundation of China (Grants No. U1832160, No. 11875071, No. 11975029, and No. 11905292).

%--------------------------------------------------------------------
\end{acknowledgments}
%--------------------------------------------------------------------

\providecommand{\href}[2]{#2}\begingroup\raggedright\endgroup

%\bibliographystyle{JHEP}
%\bibliographystyle{utphysMa}
%\bibliography{E:/yqma/work/paper/tools/JabRef/bibTex1.5}

\end{document}